\documentclass[10pt,english]{IEEEtran}
\usepackage[T1]{fontenc}
\usepackage[latin9]{inputenc}
\usepackage{float}
\usepackage{amsmath}
\usepackage{amssymb}
\usepackage{stackrel}
\usepackage{graphicx}

\usepackage[colorlinks]{hyperref}



\usepackage{babel}
\begin{document}

\title{\LARGE Fundamental Limits of Localization with Fluid Antenna Systems:\\A Fisher Information Analysis }

\author{Abdelhamid Salem, \IEEEmembership{Member, IEEE},
            Kai-Kit Wong, \IEEEmembership{Fellow, IEEE},\\
            Hyundong Shin, \IEEEmembership{Fellow, IEEE}, and
            Yangyang Zhang
\vspace{-9mm}

\thanks{(\emph{Corresponding author: Kai-Kit Wong}.)}

\thanks{The work of K. K. Wong is supported by the Engineering and Physical Sciences Research Council (EPSRC) under Grant EP/W026813/1.}
\thanks{The work of H. Shin was supported in part by the National Research Foundation of Korea (NRF) grant funded by the Korean government (MSIT) under RS-2025-00556064.}

\thanks{Abdelhamid Salem is with the Department of Electronic and Electrical Engineering, University College London, WC1E 6BT London, U.K, and also with the Department of Electronic and Electrical Engineering, Benghazi University, Benghazi, Libya. (e-mail:abdelhamid.albaraesi@gmail.com.)} 
\thanks{K. K. Wong is with the Department of Electronic and Electrical Engineering, University College London, London, United Kingdom (e-mail:kai-kit.wong@ucl.ac.uk), and also affiliated with the Department of Electronic Engineering, Kyung Hee University, Yongin-si, Gyeonggi-do 17104, Republic of Korea.}
\thanks{H. Shin is with the Department of Electronics and Information Convergence Engineering, Kyung Hee University, Yongin-si, Gyeonggi-do 17104, Republic of Korea (e-mail: hshin@khu.ac.kr).}
\thanks{Y. Zhang is with Kuang-Chi Science Limited, Hong Kong SAR, China (e-mail: yangyang.zhang@kuang-chi.org).}
}

\maketitle

\begin{abstract}
In this letter, we investigate the fundamental limits of localization in fluid antenna systems (FAS) utilizing a Fisher-information-theoretic framework. We develop a unified model to quantify the localization information extractable from time-of-arrival (ToA) and angle-of-arrival (AoA) measurements, explicitly capturing the synthetic aperture effects induced by FAS. Closed-form expressions are derived for the equivalent Fisher information matrix (EFIM) and the corresponding positioning error bound (PEB) in both user-side and base-station (BS)-side FAS configurations. Also, we propose optimal port-selection strategies based on greedy algorithms and convex relaxation to maximize the information gain under a constrained number of activated ports. Numerical results demonstrate that the proposed port-selection schemes can substantially tighten the PEB compared with random activation, thereby confirming the strong potential of FAS to enable high-precision localization. These results offer analytical insights and practical design guidelines for FAS-aided positioning in future-generation wireless networks.
\end{abstract}


\vspace{-2mm}
\section{Introduction}
\IEEEPARstart{L}{ocation}-based services are becoming an essential feature of many commercial, public service, and military wireless networks \cite{Ref1}. The ability to estimate a user's position from radio measurements, such as time-of-arrival (ToA), angle-of-arrival (AoA), and received signal strength, depends critically on the spatial geometry of the antennas involved and the richness of the propagation environment. In \cite{Ref2}, optimal precoders were derived by minimizing the Cram\'{e}r-Rao bound (CRB) of the AoA under a given uncertainty range. Moreover, \cite{Ref3} derived the CRB and proposed a novel two-stage algorithm for position and rotation angle estimation. Subsequently, in \cite{Ref4}, the receiver estimated its position based on the received reference signals, and new strategies for the power allocation to minimize the expected CRB for receiver positioning were developed. More recently, a localization error bound considering shadowing and multipath effects was derived in \cite{Ref5}.

Achieving sub-meter localization accuracy, however, often requires large physical arrays or ultra-wideband signals, both of which are difficult to realize in compact or mobile devices. To overcome these hardware constraints, the emerging concept of fluid antenna systems (FAS) has recently gained significant attention \cite{new2025a,Lu-2025,New-2026jsac}. Originally proposed by Wong {\em et al.}~in \cite{wong2020pel,wong2021fa}, FAS treats antenna as a reconfigurable physical-layer resource to empower system design with shape and position flexibility for diversity and capacity gains \cite{New2023fluid,new2023information}. Practical fluid antennas range from liquid-based antennas \cite{shen2024design,Shamim-2025}, metamaterials \cite{Liu-2025arxiv,Zhang-jsac2026}, pixel-based structures \cite{zhang2024pixel,Liu-2025iot} and others \cite[Section VI]{new2025a}. Recently, an effort was seen to use FAS under the reconfigurable intelligent surface (RIS) framework that exploits multipath for enhanced localization \cite{FAS7}.

Despite the early work in \cite{FAS7} and the potential of FAS, a theoretical characterization of FAS-enabled localization accuracy is lacking. In particular, the fundamental limits via Fisher Information Matrix (FIM) and CRB, have not been established, and the impact of port geometry, selection strategy, and base station (BS) configuration on localization performance remains unclear. This motivates the work of this letter.

In this letter, we develop an analytical framework for understanding the localization limits of FAS-assisted systems. We begin by modeling the received signal structure for both ToA and AoA estimation when a subset of antenna ports is activated from a larger set of candidate positions. Using the principles of statistical estimation theory, closed-form expressions for the equivalent FIM (EFIM) and the corresponding position error bound (PEB) are derived. The analysis is then extended to two complementary FAS deployment modalities: (i) user-side FAS, in which the mobile device hosts the reconfigurable antenna, and (ii) BS-side FAS, in which array reconfigurability is leveraged at the BS side. To enable the reconfigurability of FAS, we further formulate a port selection problem that minimizes the PEB by choosing the subset of activated ports. Two practical solutions have been developed. First, a discrete greedy algorithm that provides near-optimal performance with low computational overhead is introduced. Then, a convex relaxation approach that yields the optimum in the continuous domain, followed by rounding for discrete implementation is presented. Simulation results validate the theoretical derivations and highlight the substantial performance gains of FAS-based localization over random or static configurations. 

\vspace{-2mm}
\section{System Model}
We consider a wireless communication system consisting of $B$ BSs located at known locations, $\left\{ \mathbf{x}_{b}\in\mathbb{R}^{2},b=1,\dots,B\right\}$, and a user located at an unknown position, $\mathbf{x}\in\mathbb{R}^{2}$. Let us set the BS $b$ geometry by defining $\mathbf{d}_{b}=\mathbf{x}-\mathbf{x}_{b}$, $r_{b}=\left\Vert \mathbf{d}_{b}\right\Vert$, $\mathbf{u}_{b}=\frac{\mathbf{d}_{b}}{r_{b}}=\left[\cos\theta_{b}\;\;\;\sin\theta_{b}\right]^{T}$ where $\theta_{b}$ is the AoA. Two scenarios are considered in this work. In the first, the BS transmits signals to the user, and the user estimates its position using FAS. In the second scenario, the user transmits pilots to the BSs, which then estimate the user's position using FAS.

\vspace{-2mm}
\section{Scenario 1: User-side FAS}
In this case, the user is equipped with an $M$-port FAS,  $\mathcal{P}=\left\{ \mathbf{r}_{m},m=1,\dots,M\right\}$, where $\mathbf{r}_{m}$ is the $m$-th port location, the ports are evenly distributed along a linear dimension of length $W_{u}\lambda$ where $\lambda$ is the carrier wavelength. The BSs have fixed-position antennas (FPAs). In a localization snapshot, a subset $\mathcal{S}\subseteq\mathcal{P}$ with $|\mathcal{S}|=n_{s}$ ports are activated. Thus, the received narrowband signal at port $m\in\mathcal{S}$ from BS $b$ is
\begin{equation}
y_{b,m}(t)=\alpha_{b}e^{j2\pi f_{c}\tau_{b}}e^{j\phi_{b,m}}s\big(t-\tau_{b}\big)+n_{b,m}(t),
\end{equation}
where $f_{c}$ is the carrier frequency, $\tau_{b}=r_{b}/c$, $c$ is the speed of light, $\alpha_{b}$ is the complex channel gain, the received phase after timing alignment is $\phi_{b,m}\approx\frac{2\pi}{\lambda}\mathbf{u}_{b}^{T}\mathbf{r}_{m}$ which captures the port-dependent phase produced by the small displacement $\mathbf{r}_{m}$, i.e., the port acts like a synthetic aperture sample along direction $\mathbf{u}_{b}$, and $n_{b,m}(t)\sim\mathcal{CN}(0,\sigma^{2})$ is the additive noise.

\vspace{-2mm}
\subsection{FIM and CRB}
Generally speaking, the user position and the channel state information (CSI) can be estimated. Let the parameter vector be $\boldsymbol{\eta}=[\mathbf{x}^{T},\boldsymbol{\nu}^{T}]^{T}$, where ${\bf x}$ is the 2-D position of the user, and $\boldsymbol{\nu}$ collects the channel parameters. With circularly symmetric Gaussian noise, the complex-signal FIM is given by \cite{Ref3}
\begin{equation}
\mathbf{J}_{\boldsymbol{\eta}}=\frac{2}{\sigma^{2}}\Re\{(\partial\boldsymbol{\mu}/\partial\boldsymbol{\eta})^{T}(\partial\boldsymbol{\mu}/\partial\boldsymbol{\eta})\}.
\end{equation}
where $\boldsymbol{\mu}=\sum_{b=1}^{B}\sum_{m=1}^{n_s} \alpha_{b}e^{j2\pi f_{c}\tau_{b}}e^{j\varphi_{b,m}}$ is the mean of the received signal. To focus on user positioning, we use the Schur complement to eliminate $\boldsymbol{\nu}$. Then we can obtain the EFIM for position $\mathbf{J}_{\mathbf{x}}$. The PEB can now be defined as \cite{Ref3}
\begin{equation}
\mathrm{PEB}=\sqrt{{\rm tr}(\mathbf{J}_{\mathbf{x}}^{-1})}.
\end{equation}
where ${\rm tr}(\cdot)$ denotes the trace operation.

\vspace{-2mm}
\subsection{Information from ToA and AoA}
For BS $b$, ToA satisfies $\tau_{b}=r_{b}/c$ with gradient $\frac{\partial\tau_{b}}{\partial\mathbf{x}}=\frac{1}{c}\mathbf{u}_{b}$. The receiver estimates the arrival time with some estimation error, but the variance of this estimation error is bounded by $\sigma_{\tau,b}^{2} \geq c/2{\pi}{\beta_{\rm eff}}  \sqrt{2  {\rm SNR}}$, which is set by the signal-to-noise ratio (SNR) and the effective bandwidth ${\beta_{\rm eff}}$ \cite{loc}. Thus, the per-BS ToA EFIM contribution is given by \cite{Ref3}
\begin{equation}
\mathbf{J}_{b}^{\rm ToA}=\frac{1}{\sigma_{\tau,b}^{2}} \left({\frac{\partial\tau_{b}}{\partial\mathbf{x}}}\right)\left({\frac{\partial\tau_{b}}{\partial\mathbf{x}}}\right)^{T}=\lambda_{b}^{(\tau)}\mathbf{u}_{b}\mathbf{u}_{b}^{T},
\end{equation}
where $\lambda_{b}^{(\tau)}\triangleq\frac{1}{\sigma_{\tau,b}^{2}c^{2}}$.

Now, let $\theta_{b}=\angle(\mathbf{d}_{b})$ with gradient $\frac{\partial\theta_{b}}{\partial\mathbf{x}}=\frac{1}{r_{b}}\mathbf{u}_{b}^{\perp}$ and $\mathbf{u}_{b}^{\perp}=[0~-1~1~0]\mathbf{u}_{b}$. With the active ports in $\mathcal{S}$, the AoA estimator variance scales inversely with SNR along direction $\mathbf{u}_{b}^{\perp}$, and the number of the active ports $n_{s}$. For port $m$, the received phase (after timing alignment) is approximately
\begin{equation}
\phi_{b,m}\approx\frac{2\pi}{\lambda}\mathbf{u}_{b}^{T}\mathbf{r}_{m},
\end{equation}
and its sensitivity to $\theta_{b}$, can be measured by
\begin{equation}
\frac{\partial\phi_{b,m}}{\partial\theta_{b}}=\frac{2\pi}{\lambda}\frac{\partial\mathbf{u}_{b}^{T}}{\partial\theta_{b}}\mathbf{r}_{m}=\frac{2\pi}{\lambda}(\mathbf{u}_{b}^{\perp})^{T}\mathbf{r}_{m}.
\end{equation}
If the per-port phase noise is independent and identically distributed (i.i.d.) with variance $\sigma_{\phi}^{2}$, the scalar FIM for $\theta_{b}$ is
\begin{equation}
J_{\theta\theta}^{b}=\frac{1}{\sigma_{\phi}^{2}}\sum_{m\in\mathcal{S}}\left(\frac{\partial\phi_{b,m}}{\partial\theta_{b}}\right)^{2}=\frac{(2\pi/\lambda)^{2}}{\sigma_{\phi}^{2}}\sum_{m\in\mathcal{S}}\Big((\mathbf{u}_{b}^{\perp})^{T}\mathbf{r}_{m}\Big)^{2}.
\end{equation}
Hence, the AoA variance can be found as
\begin{align}
\sigma_{\theta,b}^{2}=\frac{1}{J_{\theta\theta}^{b}}&=\frac{\sigma_{\phi}^{2}}{(2\pi/\lambda)^{2}\sum_{m\in\mathcal{S}}((\mathbf{u}_{b}^{\perp})^{T}\mathbf{r}_{m})^{2}},\notag\\
&=\frac{\sigma_{\phi}^{2}}{\kappa_{b}\sum_{m\in\mathcal{S}}(\mathbf{u}_{b}^{\perp T}\mathbf{r}_{m})^{2}},
\end{align}
where $\kappa_{b}$ is the wavelength and calibration constant, e.g., $\kappa_{b}=(2\pi/\lambda)^{2}$. Thus, the per-BS AoA EFIM contribution is 
\begin{equation}
\mathbf{J}_{b}^{\rm AoA}(\mathcal{S})=\frac{1}{\sigma_{\theta,b}^{2}} \left({\frac{\partial\theta_{b}}{\partial\mathbf{x}}}\right)\left({\frac{\partial\theta_{b}}{\partial\mathbf{x}}}\right)^{T}=\lambda_{b}^{(\theta)}(\mathcal{S})\frac{\mathbf{u}_{b}^{\perp}(\mathbf{u}_{b}^{\perp})^{T}}{r_{b}^{2}},
\end{equation}
where $\lambda_{b}^{(\theta)}(\mathcal{S})\triangleq\frac{1}{\sigma_{\theta,b}^{2}}\propto\kappa_{b}\sum_{m\in\mathcal{S}}(\mathbf{u}_{b}^{\perp T}\mathbf{r}_{m})^{2}$.

The sum FIM is therefore given by 
\begin{equation}
\mathbf{J}_{\mathbf{x}}(\mathcal{S})=\sum_{b=1}^{B}\left[\lambda_{b}^{(\tau)}\mathbf{u}_{b}\mathbf{u}_{b}^{T}+\lambda_{b}^{(\theta)}(\mathcal{S})\frac{\mathbf{u}_{b}^{\perp}(\mathbf{u}_{b}^{\perp})^{T}}{r_{b}^{2}}\right]
\end{equation}
and 
\begin{equation}
\mathrm{PEB}=\sqrt{{\rm tr}\left[\mathbf{J}_{\mathbf{x}}^{-1}(\mathcal{S})\right]}.
\end{equation}

As we can observe from the previous expressions, ToA adds information along ($\mathbf{u}_{b})$, while AoA adds information orthogonal to ($\mathbf{u}_{b}$), with strength controlled by how the selected ports span the port perpendicular to the BS direction.

\vspace{-2mm}
\section{Scenario 2: BS-side FAS}
Now, each BS uses a fluid antenna. For BS $b$ with $M$ ports, we have $\mathcal{P}_{b}=\left\{ \mathbf{r}_{b,m},m=1,\dots,M\right\}$, where $\mathbf{r}_{b,m}$ is the $m$-th port location at BS $b$. The ports are evenly distributed along a linear dimension of length $W_{b}\lambda$, while a user at an unknown location $\mathbf{x}$ is equipped with a FPA. In a localization snapshot, a subset $\mathcal{S}_{b}\subseteq\mathcal{P}_{b}$ with $|\mathcal{S}_{b}|=n_{s}$ ports are activated. 

\vspace{-2mm}
\subsection{Information from ToA and AoA}
The user transmits pilots to the associated BS, then the BS estimates the AoA $\theta_{b}$ from the spatial phase slopes across its ports. 
Similarly as in the previous scenario, we can obtain  the AoA variance at BS $b$ as 
\begin{equation}
\sigma_{\theta,b}^{2}=\frac{\sigma_{\phi,b}^{2}}{(2\pi/\lambda)^{2}\sum_{m\in\mathcal{S}_{b}}((\mathbf{u}_{b}^{\perp})^{T}\mathbf{r}_{b,m})^{2}},
\end{equation}
where $\sigma_{\phi,b}^{2}$ denotes the per-port phase noise variance. Using the scalar-measurement lemma, the angular term driven by the BS-port geometry via $\sigma_{\theta,b}^{2}$ is given by
\begin{equation}
\mathbf{J}_{b}^{\rm AoA}=\lambda_{b}^{(\theta)} (\mathcal{S}_{b})\frac{\mathbf{u}_{b}^{\perp}(\mathbf{u}_{b}^{\perp})^{T}}{r_{b}^{2}},
\end{equation}
where $\lambda_{b}^{(\theta)}(\mathcal{S}_{b})=\frac{1}{\sigma_{\theta,b}^{2}}=\Big(\tfrac{2\pi}{\lambda}\Big)^{2}\tfrac{1}{\sigma_{\phi,b}^{2}}\sum_{m\in\mathcal{S}_{b}}\big((\mathbf{u}_{b}^{\perp})^{T}\mathbf{r}_{b,m}\big)^{2}$.

Also, the ToA EFIM can be written as 
\begin{equation}
\mathbf{J}_{b}^{\rm ToA}=\lambda_{b}^{(\tau)}\mathbf{u}_{b}\mathbf{u}_{b}^{T},~\mbox{where }\lambda_{b}^{(\tau)}=\frac{1}{\sigma_{\tau,b}^{2}c^{2}}.
\end{equation}

As a result, overall, the per-BS EFIM and the network EFIM are given, respectively, by 
\begin{equation}
\mathbf{J}_{\mathbf{x},b}=\lambda_{b}^{(\tau)}\mathbf{u}_{b}\mathbf{u}_{b}^{T}+\lambda_{b}^{(\theta)}(\mathcal{S}_{b})\frac{\mathbf{u}_{b}^{\perp}(\mathbf{u}_{b}^{\perp})^{T}}{r_{b}^{2}},
\end{equation}
and
\begin{equation}
\mathbf{J}_{\mathbf{x}}=\sum_{b=1}^{B}\mathbf{J}_{\mathbf{x},b}.
\end{equation}

This structure is identical to the user-FAS case, the difference lies in how $\lambda_{b}^{(\theta)}$ is generated, the BS-port geometry and the BS-side SNR, instead of the user-port geometry.

\vspace{-2mm}
\section{Port Selection}
In this section, we select the optimal $n_{s}$ ports to tighten the CRB. To do so, we employ a classic optimal design criteria to maximize $\log\det\mathbf{J}_{\mathbf{x}}$ which minimizes PEB. Maximizing the objective function $\log\det\mathbf{J}_{\mathbf{x}}$ is a standard approach in optimal design used to minimize the PEB or the more general CRB on parameter estimation \cite{logdet}.

\vspace{-2mm}
\subsection{Discrete-Greedy Scheme}
Define each port's AoA information kernel toward BS $b$ as
\begin{equation}
\mathbf{Q}_{b,m}\triangleq\gamma_{b,m}\frac{\mathbf{u}_{b}^{\perp}(\mathbf{u}_{b}^{\perp})^{T}}{r_{b}^{2}},
\end{equation}
where $\gamma_{b,m}=\lambda_{b}^{(\theta)}(\mathcal{S})$ for Scenario 1 and $\gamma_{b,m}=\lambda_{b}^{(\theta)}(\mathcal{S}_{b})$ for Scenario 2. Letting $\mathbf{J}_{0}=\sum_{b}\lambda_{b}^{(\tau)}\mathbf{u}_{b}\mathbf{u}_{b}^{T}$ (i.e., the ToA part, independent of port choice), we can then write 
\begin{equation}
\mathbf{J}_{\mathbf{x}}=\mathbf{J}_{0}+\sum_{b}\sum_{m}\mathbf{Q}_{b,m}.
\end{equation}

A greedy scheme adds the port with the largest marginal log-det gain to the set
\begin{multline}
m^{\star}=\\
\arg\max_{m\in\mathcal{P}\setminus\mathcal{P}_{b}}\left[\log\det\left(\mathbf{J}_{0}+\sum_{b}\mathbf{Q}_{b,m}\right)-\log\det(\mathbf{J}_{0})\right].
\end{multline}
We then update EFIM ($\mathbf{J}\leftarrow\mathbf{J}+\sum_{b}\mathbf{Q}_{b,m}$), and the selected ports ($\mathcal{S}\leftarrow\mathcal{S}\cup m^{\star}$). These procedures are iteratively repeated until the optimal ports, $n_{s}$, are identified. This is the standard lazy-greedy scheme. Note that $\log\det$ is monotone and close to sub-modular in such additive EFIM settings, achieving near-optimality with $O(Mn_{s})$ determinant updates.

\vspace{-2mm}
\subsection{Convex Continuous Relaxation Scheme}
In this method, we first introduce selection weights $x_{m}\in[0,1]$ with $\sum_{m}x_{m}=n_{s}$. Then we can write
\begin{equation}
\mathbf{J}(x)=\mathbf{J}_{0}+\sum_{b}\sum_{m} x_{m}\mathbf{Q}_{b,m}.
\end{equation}

Note that the function $-\log\det(\cdot)$ is convex over symmetric positive definite (SPD) matrices and $\mathbf{J}(x)$ is affine in $x$. Thus we can formulate the problem as
\begin{equation}
\min_{x\in[0,1]^{M}}-\log\det \mathbf{J}(x)~\mbox{s.t.}~\sum_{m=1}^{M}x_{m}=n_{s}.
\end{equation}
This problem is a convex program, which can be solved with standard solvers (interior-point algorithm). After that we round the result by taking the top $n_{s}$ entries of $x^{\star}$.

\vspace{-2mm}
\section{Numerical Results}
In this section, we evaluate the localization performance of the proposed FAS-assisted architecture. The BSs are placed symmetrically around the origin, each located $50~{\rm m}$ from the center, while the user's position is assumed to be near the center. The main parameters are: carrier frequency $f_{c}=3~{\rm GHz}$, the number of BSs $B=4$, the number of FAS ports $M=60$, the number of selected ports $n_{s}=10$, the FAS size at the user $l_{u}=0.5\lambda$ and the FAS size at the BSs $l_{b}=2\lambda$. 

\begin{figure}[]
\centering
\includegraphics[width=.78\columnwidth]{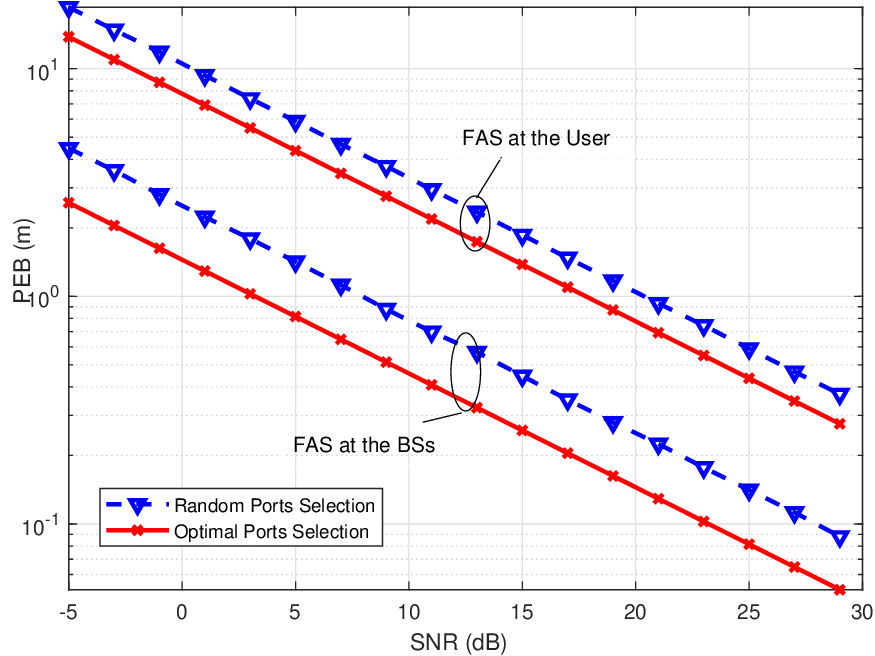}
\caption{PEB vs SNR for different FAS configurations.}\label{fig_1}
\vspace{-2mm}
\end{figure}

Fig.~\ref{fig_1} shows the PEB versus the SNR for the two cases, FAS implemented at the user side and at the BSs side. Two port-selection strategies, random and optimal schemes, are shown. As SNR increases, both configurations exhibit a monotonic reduction in PEB, consistent with the increase in Fisher information. However, the BS-side FAS achieves significantly lower PEB across the entire SNR range owing to its larger aperture, which improves the angular estimation precision. Optimal port selection further tightens the PEB compared to random activation, demonstrating that intelligent control of the active ports yields substantial localization gains.

\begin{figure}[]
\centering
\includegraphics[width=.76\columnwidth]{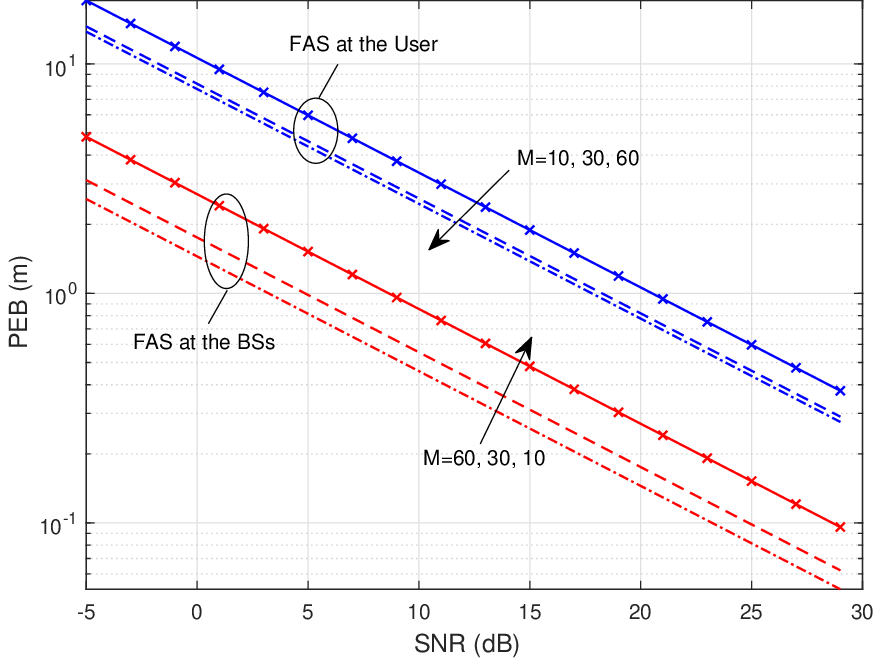}
\caption{PEB vs SNR for different FAS configurations and $M$ values.}\label{fig_2}
\vspace{-2mm}
\end{figure}

Fig.~\ref{fig_2} investigates how the number of  ports $M$ influences localization accuracy when the number of active ports is fixed $n_{s}=10$. Three different configurations, $M=10,30,60$ are compared for both the user-FAS and BS-FAS setups. As $M$ increases, the PEB decreases because a denser set of candidate positions allows the proposed algorithms to form more favorable synthetic ports with larger effective spatial spread. This figure confirms that denser FAS ports enhances angular diversity and improves the achievable resolution, especially at moderate-to-high SNRs. Again, BS-side FAS consistently outperforms user-side FAS due to its superior aperture.

\begin{figure}[]
\centering
\includegraphics[width=.76\columnwidth]{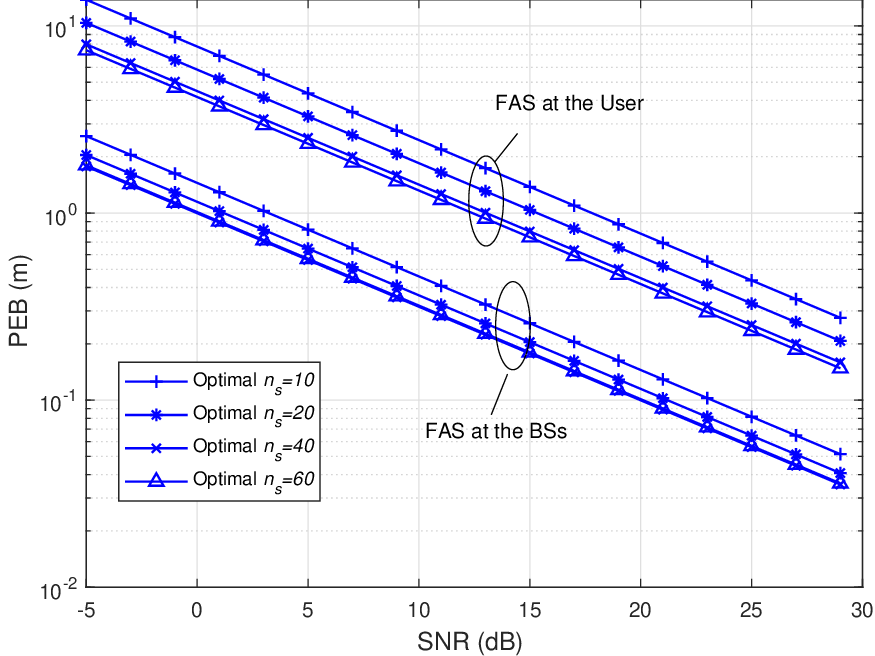}
\caption{PEB vs SNR different FAS configurations and $n_{s}$ values.}\label{fig_3}
\vspace{-2mm}
\end{figure}

Fig.~\ref{fig_3} examines the effect of varying the number of active ports $n_{s}=10,20,40,60$ under the optimal selection strategy when $M=60$. Increasing $n_{s}$ yields a clear improvement in localization accuracy because each additional active port contributes an independent angular measurement, effectively enlarging the equivalent Fisher information. The improvement follows a diminishing-returns trend\textemdash most of the gain occurs when moving from very few to moderate numbers of ports, after which the benefit saturates. The BS-side FAS consistently achieves smaller PEB values, reaffirming its advantage in geometry. At high SNRs and large $n_{s}$, sub-decimeter accuracy becomes attainable even with compact fluid ports.

\vspace{-2mm}
\section{Conclusions}
In this letter, we investigated the localization performance limits of wireless systems enhanced by FAS deployed either at the user side or at the BS. A unified Fisher information-based framework was developed to analyze the achievable PEB. Closed-form expressions for the EFIM were also derived under narrowband channel assumptions, incorporating both AoA and ToA information. The impact of FAS geometry, SNR, and port-selection strategies were evaluated. Numerical results demonstrated that optimal port selection significantly improves localization accuracy compared to random activation by maximizing the EFIM determinant. The results also confirmed that increasing either the number of candidate ports or the number of active ports enhances localization performance, although with diminishing returns at high densities.

\bibliographystyle{IEEEtran}

\end{document}